\documentclass[journal,twoside,web]{ieeecolor}
\usepackage{tmi}
\usepackage{cite}
\usepackage{amsmath,amssymb,amsfonts,mathtools}
\usepackage{algorithmic}
\usepackage{graphicx}
\usepackage{textcomp}
\usepackage{setspace}
\usepackage{svg}
\usepackage{algorithm}
\usepackage{flushend}
\usepackage{siunitx}
\usepackage{hyperref}
\usepackage{siunitx}
\usepackage{wrapfig}
\usepackage{placeins}
\usepackage{orcidlink}

\def\BibTeX{{\rm B\kern-.05em{\sc i\kern-.025em b}\kern-.08em
    T\kern-.1667em\lower.7ex\hbox{E}\kern-.125emX}}
\newcommand{\bx}{{\mathbf x}}
\newcommand{\by}{{\mathbf y}}
\newcommand{\bh}{{\mathbf h}}
\newcommand{\bz}{{\mathbf z}}

\newcommand{\bI}{{\mathbf I}}
\newcommand{\bR}{{\mathbb R}}

\newcommand{\norm}[1]{\left\lVert#1\right\rVert}

\DeclareMathOperator*{\argmin}{arg\,min}
\DeclareMathOperator{\sign}{sgn}

\newcommand{\algocomment}[1]{\textcolor{mblue}{\footnotesize{\texttt{\textbf{// #1}}}}}

\newcommand{\TSW}[1]{{\textcolor{black}{#1}}}
\newcommand{\TSWR}[1]{{\textcolor{black}{#1}}}

\begin{document}
\title{Dehazing Ultrasound using Diffusion Models}
\author{Tristan S.W. Stevens\orcidlink{0000-0002-8563-5931}, \IEEEmembership{Graduate Student Member, IEEE}, Faik C. Meral, Jason Yu, Iason Z. Apostolakis,\\Jean-Luc Robert and Ruud J.G. van Sloun\orcidlink{0000-0003-2845-0495}, \IEEEmembership{Member, IEEE}
\thanks{
Manuscript received 11 December 2023; revised 25 January 2024;
accepted 2 February 2024. Date of publication 7 February 2024; date
of current version 24 October 2024. This work was performed within the IMPULSE framework of the Eindhoven MedTech Innovation Center (e/MTIC, incorporating Eindhoven University of Technology and Philips Research), including a PPS supplement from the Dutch Ministry of Economic Affairs and Climate Policy.\\
\emph{(Corresponding author: Tristan S. W. Stevens.)}\\
\indent This work involved human subjects or animals in its research. Approval
of all ethical and experimental procedures and protocols was granted by
the Philips Internal Committee for Biomedical Experiments (ICBE) under
application no. ICBE-S-000092.\\
\indent Tristan S. W. Stevens and Ruud J. G. van Sloun are with the Electrical Engineering Department, Eindhoven University of Technology,
5612 AZ Eindhoven, The Netherlands (e-mail: t.s.w.stevens@tue.nl;
r.j.g.v.sloun@tue.nl).\\
\indent Faik C. Meral, Jason Yu, Iason Z. Apostolakis, and Jean-Luc Robert are with Philips Research, Cambridge, MA 02141 USA (e-mail: can.meral@philips.com; jason.yu\_1@philips.com; iasonzacharias.
apostolakis@philips.com; jean-luc.robert@philips.com).\\
\indent Digital Object Identifier 10.1109/TMI.2024.3363460}
}

\maketitle

\begin{abstract}
Echocardiography has been a prominent tool for the diagnosis of cardiac disease. However, these diagnoses can be heavily impeded by poor image quality. Acoustic clutter emerges due to multipath reflections imposed by layers of skin, subcutaneous fat, and intercostal muscle between the transducer and heart. As a result, haze and other noise artifacts pose a real challenge to cardiac ultrasound imaging. In many cases, especially with difficult-to-image patients such as patients with obesity, a diagnosis from B-Mode ultrasound imaging is effectively rendered unusable, forcing sonographers to resort to contrast-enhanced ultrasound examinations or refer patients to other imaging modalities. Tissue harmonic imaging has been a popular approach to combat haze, but in severe cases is still heavily impacted by haze. Alternatively, denoising algorithms are typically unable to remove highly structured and correlated noise, such as haze. It remains a challenge to accurately describe the statistical properties of structured haze, and develop an inference method to subsequently remove it. Diffusion models have emerged as powerful generative models and have shown their effectiveness in a variety of inverse problems. In this work, we present a joint posterior sampling framework that combines two separate diffusion models to model the distribution of both clean ultrasound and haze in an unsupervised manner. Furthermore, we demonstrate techniques for effectively training diffusion models on radio-frequency ultrasound data and highlight the advantages over image data. Experiments on both \emph{in-vitro} and \emph{in-vivo} cardiac datasets show that the proposed dehazing method effectively removes haze while preserving signals from weakly reflected tissue.
\end{abstract}

\newpage

\section{Introduction}
\label{sec:introduction}

\IEEEPARstart{C}{ardiovascular} ultrasound imaging or echocardiography has been one of the most important advances in cardiac imaging due to its real-time nature and cost-effectiveness. Cardiac ultrasound allows cardiologists to review the basic functioning of the heart and detect abnormalities which are prominent markers for cardiovascular diseases that make up 31\% of all global deaths \cite{daveRecentTechnologicalAdvancements2018}. Unfortunately, transthoracic echocardiograms are subject to a range of different noise sources that clutter the image and limit interpretability. Acoustic clutter, i.e. any unwanted echoes that degrade the image quality, is caused by phenomena such as aberration and reverberation. A specific type of noise artifact, known as haze, emerges due to multipath reflections imposed by layers of skin, subcutaneous fat, and intercostal muscle that acoustic energy has to traverse before reaching the heart \cite{fatemiStudyingOriginReverberation2019}. Moreover, specular reflection of acoustic signals at the ribs causes deflection of the propagation path. Combined, these multipath signals culminate into a white diffuse haze over the image, most dominantly present in the near-field. This effect is especially significant in technically difficult-to-image patients such as those with obesity or dense muscle structures, often leading to nondiagnostic examinations \cite{uppotImpactObesityRadiology2007}.

Image quality is vital in cardiac ultrasound imaging, as analysis of the structure and function of the heart relies on careful interpretation of the brightness mode (B-mode) sequence, both by the clinician and post-formation image processing techniques \cite{sassaroliImageQualityEvaluation2019, chenDeepLearningCardiac2020}. Hence, lower image quality can limit functional diagnosis as both visual assessment and quantitative analysis become challenging with hazy ultrasound data. Tissue harmonic imaging was introduced to combat deteriorating image quality in technically challenging patients \cite{tranquartClinicalUseUltrasound1999}, \cite{wellsUltrasoundImaging2006}. Compared to fundamental imaging, harmonics contain minimal clutter and noise. Multipath and distorted scatterers are much weaker in energy and therefore generate fewer harmonics leading to generally improved signal-to-noise and contrast-to-noise ratios. However, harmonic imaging comes at the cost of reduced penetration depth and image frame rate (due to the additional pulse inversion transmits). Furthermore, in many severe cases, the haze artifact persists even with harmonic imaging. This results in costly repeat examinations, either with contrast-enhanced ultrasound or referrals to other image modalities (MRI / CT). This highlights the need for improved image quality in cardiac ultrasound. 

Deep generative modeling has gained traction in the medical field and has successfully been used for image reconstruction in MRI, CT, and ultrasound \cite{yangDAGANDeepDeAliasing2018}, \cite{youCTSuperResolutionGAN2020}, \cite{songSolvingInverseProblems2022}, \cite{vandeschaftUltrasoundSpeckleSuppression2021}, \cite{chungScorebasedDiffusionModels2022}. These data-driven methods are more powerful compared to classical methods, as they can accurately learn the natural signal manifold and do not rely on basic assumptions such as signal sparsity or hand-crafted basis functions \cite{vanslounDeepLearningUltrasound2020}. More recently, diffusion models have proven to be able to accurately model complex and high-dimensional data distributions. In contrast to \emph{discriminative} methods, diffusion models are a type of \emph{generative} models and can be trained on unlabeled and unpaired datasets which are often costly to come by in the medical domain. Furthermore, deep generative models are more flexibly applicable as the training part is agnostic to the reconstruction task. This also relates to improved generalizability across datasets, patients, and even modalities, which discriminative methods often struggle with \cite{zhangDeepStableLearning2021}, \cite{chatterjeeGeneralizationMysteryDeep2022}. As diffusion models indirectly parameterize the data distribution through its gradient, they come with some advantages compared to alternative generative models. Diffusion models are easier to train relative to adversarial approaches such as GANs, which suffer from modal collapse, and have no restrictions on the neural architecture unlike normalizing flows \TSW{\cite{kingma2018glow}}. 

In this work, we adopt advances in conditional sampling with diffusion models which seek to sample from a learned data distribution given some measurement. More specifically, we explicitly model both clean tissue and haze using separate score-based networks. This allows us to capture the highly structured and spatially correlated nature of haze, unlike many methods that have a basic assumption on the noise distribution, often to assure tractability. Additionally, we perform the entire dehazing process including training of the priors in the radio-frequency (RF) domain. This allows us to exploit phase information still present in pre-envelope detected ultrasound data, as well as linearly separate signal and haze sources early in the signal chain. To test the proposed dehazing method we perform tests in both a controlled phantom experiment and on cardiac \emph{in-vivo} datasets. With a classical denoising algorithm as well as a supervised learning model as the baseline methods, we show a consistent improvement in generalized contrast-to-noise ratio (gCNR), whilst preserving more low-level tissue in the process.

\section{Related Work}

Fatemi~\emph{et al.}, investigate the cause for reverberations leading to artifacts commonly seen in cardiac ultrasound imaging \cite{fatemiStudyingOriginReverberation2019}. They found several possible scenarios that generate strong reverberation clutter, including multipath reflections caused by tissue layers between the transducer and the heart, specular reflections at the ribs, and blocking of the ultrasound beam by the lungs. Furthermore, in their effort to compare fundamental and harmonic imaging, haze is often reduced using the latter technique. Still, haze persists in many of the examples showing the need for improved techniques. 

Post-processing methods for clutter mitigation, using methods such as block-matching and 3D filtering algorithm (BM3D) \cite{dabovImageDenoisingBlockmatching2006}, \cite{santosUltrasoundImageDespeckling2017} or wavelet-based denoising \cite{yueNonlinearMultiscaleWavelet2006}, are typically based on computing local statistics and impose basic assumptions on the noise distribution. Furthermore, image-based denoising methods are often paired with inherent speckle reduction or smoothing. Speckle alone is generally not the cause for limited visibility of important cardiac features and in many cases advantageous for the preservation of finer detail and methods relying on speckle tracking \cite{geyerAssessmentMyocardialMechanics2010}. 

Sjoerdsma~\emph{et al.}, introduce a near-field clutter filter that operates in the spatial frequency domain \cite{sjoerdsmaSpatialNearFieldClutter2021}. Their method preserves speckle and addresses the haze in the near field using post-envelope detected image data. However, their method was optimized for strain computation and suppresses tissue at the myocardial boundary, obfuscating finer low-level structures.

Methods based on temporal decompositions such as principal component analysis (PCA) and singular value decomposition (SVD)  \cite{turekClutterMitigationEchocardiography2015}, \cite{solomonDeepUnfoldedRobust2020}, \cite{wildeboerBlindSourceSeparation2020} leverage the stationarity of clutter with respect to more rapidly moving tissue. However, there are cases in the cardiac cycle where this assumption does not hold such as for tissue at end-diastole or generally more stationary myocardial regions. Hence, in these instances, the lack of sophistication of the temporal prior inhibits these methods to separate tissue from the clutter.

Deep learning has become a popular tool for ultrasound image processing 
\cite{vandeschaftUltrasoundSpeckleSuppression2021}, \cite{vanslounDeepLearningUltrasound2020}, \cite{liuDeepLearningMedical2019}, \cite{vignonResolutionImprovementFully2020}, \cite{stevensAcceleratedIntravascularUltrasound2022}, \cite{chennakeshavaDeepProximalUnfolding2022}, \cite{luijtenUltrasoundSignalProcessing2023}.
More specifically, Jahren~\emph{et al.} present a supervised learning approach to suppress reverberation clutter in cardiac ultrasound using beamformed data \cite{jahrenReverberationSuppressionEchocardiography2023}. In contrast to the proposed work, their method is discriminative, lacking the aforementioned advantages of generative models such as diffusion models.

Despite the recent success of diffusion models, their relevance to ultrasound has seen very limited exploration \cite{asgariandehkordiDeepUltrasoundDenoising2023}, \cite{stojanovskiEchoNoiseSynthetic2023}. Nonetheless, diffusion models for image reconstruction have been successfully applied to other modalities in the medical field, such as CT and MRI \cite{songSolvingInverseProblems2022}, \cite{chungScorebasedDiffusionModels2022}, \cite{kazerouniDiffusionModelsMedical2023}.

The remainder of the paper is organized as follows. In Section~\ref{sec:methods}, we provide a formulation of the dehazing process (\ref{sec:ultrasound}), give background on score-based diffusion models (\ref{sec:diffusion}), and introduce our ultrasound dehazing method (\ref{sec:dehazing}). Our main contributions involve methods for efficiently learning ultrasound priors using diffusion models and combining them to perform posterior sampling for dehazing. The results on both phantom and \emph{in-vivo} experiments are given in Section~\ref{sec:results}. Lastly, we discuss the results in Section~\ref{sec:discussion} and derive conclusions in Section~\ref{sec:conclusions}.
\begin{figure}
    \centering
    \includegraphics[scale=1.3]{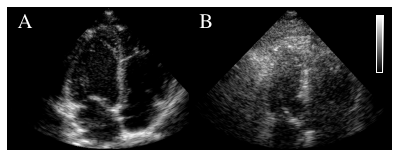}
    \caption{\hspace{-0.5em}Echocardiograms of easier (A) and difficult-to-image patients (B).}
    \label{fig:haze_comparison}
    \vspace{-0.5cm}
\end{figure}

\section{Methods}
\label{sec:methods}
In this section, we start by providing the reader insight into the ultrasound acquisition process and the specific formulation used leading to a probabilistic framework for dehazing. We then turn our attention to diffusion models and how we can efficiently learn ultrasound priors in the RF domain for the dehazing task.

\subsection{Ultrasound Image Formation}
\label{sec:ultrasound}
\begin{figure}
    \centering
    \includegraphics[scale=0.8]{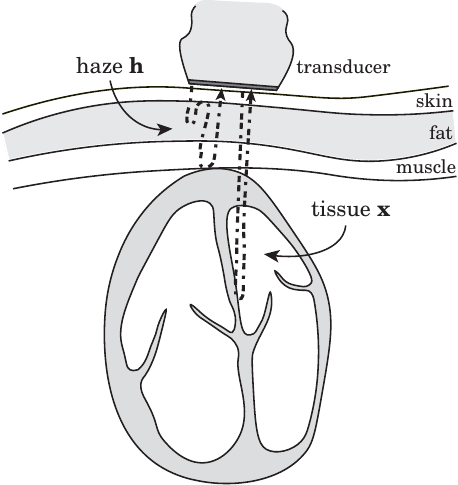}
    \caption{Simplified schematic of a phased array transducer performing an echocardiogram that shows a possible scenario of how the ultrasound energy can traverse through the skin and underlying tissue structures. We decompose the receiving ultrasound energy into haze signals $\bh$ (multipath reverberation) and tissue signals $\bx$.}
    \label{fig:haze_schematic}
    \vspace{-0.5cm}
\end{figure}

The beamformed ultrasound signal can be considered a coherent summation of individual ultrasonic backscattered echoes. These \emph{scatterers} originate from small structures in the tissue. As a result, we can group the signals reflected by the tissue $\bx$ and all multipath haze signals $\bh$, leading to the following additive model:
\begin{align}
    \by_{\text{RF}} = \bx_{\text{RF}} + \bh_{\text{RF}},
    \label{eq:inverse_model}
\end{align}
where $\by$ is the ultrasound measurement. For clarity, if necessary, subscripts are added to denote quantities as beamformed radio-frequency (RF) data. A schematic of the measurement process is shown in Fig.~\ref{fig:haze_schematic}. The ultrasound image, also known as the B-mode image, is obtained by envelope detection and log compression of the RF data. We here pose the dehazing problem as a source separation task, where we would like to retrieve the clean ultrasound signal $\bx$ from the measurement $\by$. This inverse problem is generally ill-posed, as there are many possible solutions that satisfy \eqref{eq:inverse_model}. Following a probabilistic approach, sampling from the posterior distribution $p_{X, H}(\bx, \bh|\by)$ allows us to find optimal $\hat{\bx}$ and $\hat{\bh}$ given the measurement $\by$ and prior knowledge. To compute the posterior, we can factorize it using Bayes' rule as follows:
\begin{align}
    (\bx, \bh) &\sim p_{X, H}(\bx, \bh|\by) \propto p_{Y|X,H}(\by|\bx, \bh) \cdot p_X(\bx) \cdot p_H(\bh),
    \label{eq:bayes}
\end{align}
where $p_{Y|X,H}(\by|\bx, \bh)$ is the likelihood according to our measurement model in \eqref{eq:inverse_model} and $p_X(\bx)$ and $p_H(\bh)$ are prior distributions for the clean ultrasound signal and haze, respectively. Note that, unlike many (denoising) methods, we consider the noise distribution to be any arbitrarily complex distribution and thus not necessarily Gaussian distributed, i.e. $p_H(\bh) \neq \mathcal{N}$. This is an important distinction, given the fact that haze is a noticeably more structured and spatially correlated type of noise.
\begin{figure*}[t]
    \centering
    \includegraphics[scale=0.2]{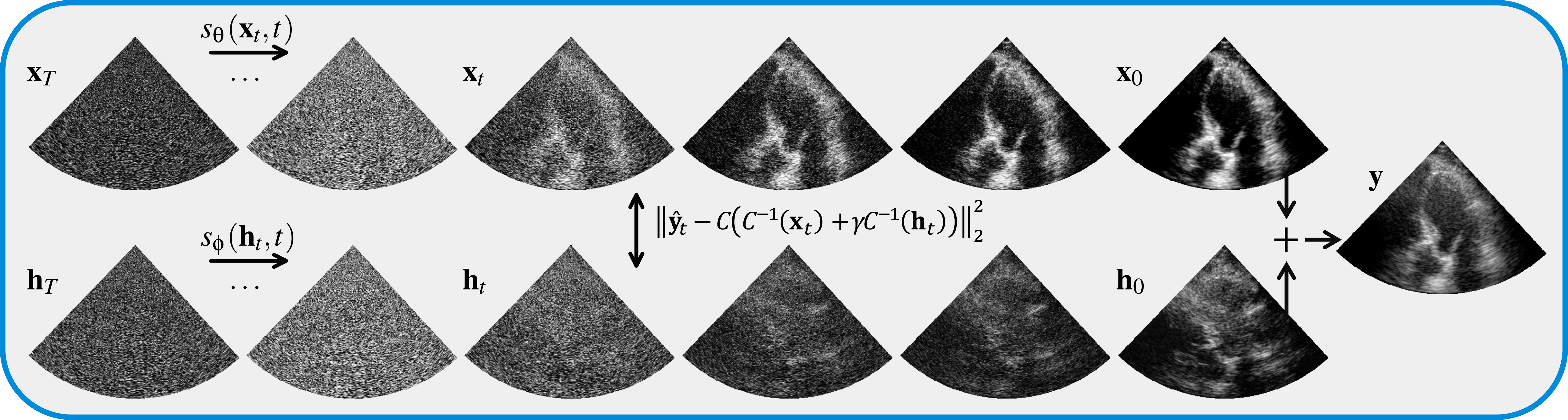}
    \caption{Dehazing diffusion process, where the reverse diffusion trajectory is displayed from left to right for both signal (top) and haze (bottom) in parallel. \TSWR{Each reverse diffusion step is carried out by a denoising U-Net that models the score $s(\cdot)$ related to either signal or haze.} During each step of the posterior sampling process, data consistency is promoted through the measurement model $\by_{\text{RF}} = \bx_{\text{RF}} + \bh_{\text{RF}}$.}
    \label{fig:diffusion_dehazing}
\end{figure*}
\subsection{Score-Based Diffusion Models}
\label{sec:diffusion}
In this work, we opt to learn the prior distributions of clean ultrasound RF data and haze using score-based diffusion models \cite{hoDenoisingDiffusionProbabilistic2020, songScoreBasedGenerativeModeling2021, karrasElucidatingDesignSpace2022}. These deep generative models aim to reverse a diffusion process, which corrupts clean data $\bx_0 \sim p(\bx_0)\equiv p(\bx)$ to some predefined base distribution $p_1(\bx)\approx\pi(\bx)$ through a sequence of Gaussian perturbations indexed by time $t\in[0, 1]$. The forward diffusion process can be described using a stochastic differential equation (SDE) as follows:
\begin{equation}
    \mathrm{d}\bx_t = f(t)\bx_t\mathrm{d}t + g(t) \mathrm{d}\mathbf{w},
\end{equation}
where $\mathbf{w}\in\bR^d$ is a standard Wiener process, $f(t): [0, 1]\rightarrow\bR$ and $g(t): [0, 1]\rightarrow\bR$ are the drift and diffusion coefficients that account for the deterministic and stochastic parts of the diffusion process, respectively. 

Naturally, we are interested in reversing this diffusion process, so that we can sample from $\bx_0 \sim p(\bx_0)$. The reverse diffusion process is also a diffusion process given by the reverse-time SDE \cite{songScoreBasedGenerativeModeling2021}:
\begin{equation}
    \mathrm{d}\bx_t = 
    \big[
        f(t)\bx_t - g(t)^2\underbrace{\nabla_{\bx_t} \log{p(\bx_t)}}_{\text{score}}
    \big] \mathrm{d}t + g(t)\mathrm{d}\Bar{\mathbf{w}}_t,
    \label{eq:reverse_diff}
\end{equation}
where $\Bar{\mathbf{w}}_t$ is the standard Wiener process in the reverse direction. The gradient of the log-likelihood of the data with respect to itself, a.k.a. the \textit{score function}, arises from the reverse-time SDE. The score function is a gradient field pointing back to the data manifold and can intuitively be used to guide a random sample from the base distribution $\pi(\bx)$ to the desired data distribution $p(\bx_0)$. The score function itself can be learned by a neural network $s_\theta(\bx_t, t)$ parameterized by weights $\theta$. These weights can be optimized using score-matching techniques, such as denoising score-matching (DSM) \cite{vincentConnectionScoreMatching2011}. The objective of DSM is given by:
\begin{align}
    \theta^* = \argmin_\theta \mathbb{E}_{t\sim U[0, 1]}
    \biggl\{
        \mathbb{E}_{(\bx, \bx_t)\sim p(\bx)q(\bx_t|\bx)}  \nonumber \\
        \left[ 
            \norm{s_\theta(\bx_t, t) - \nabla_{\bx_t}\log q(\bx_t|\bx)}_2^2
        \right]
    \biggr\},
    \label{eq:score_matching}
\end{align}
where $q(\bx_t|\bx_0)\sim\mathcal{N}$ is the perturbation kernel of the diffusion process which is also Gaussian due to the properties of the SDE. Given a sufficiently large dataset $\mathcal{X} = \left\{\bx_0^{(1)}, \bx_0^{(2)}, \ldots,  \bx_0^{(|\mathcal{X}|)}\right\}\sim p(\bx_0)$ and enough model capacity, DSM enables approximating the true score $s_\theta(\bx_t, t)\simeq \nabla_{\bx_t}\log p(\bx_t)$. During inference, we can substitute our learned score in \eqref{eq:reverse_diff} and discretize the reverse-time diffusion process into a sequence of time steps $\left\{0=t_0, t_1, \ldots, t_T=1 \right\}$. Numerical samplers such as the Euler-Maruyama method \cite{songScoreBasedGenerativeModeling2021} can be used to solve the discretized reverse-time SDE in an iterative fashion using the trained score network. The update rule for this inference procedure is given by:
\begin{equation}
\bx_{t - \Delta t} \leftarrow \bx_t + [f(t)\bx_t - g^2(t) s_\theta(\bx_t)]\Delta t + g(t) \sqrt{\vert\Delta t\vert}\bz,
\end{equation}
with $\bz \sim \mathcal{N}\left(0, \bI\right)$. During the reverse-time diffusion process, the solution is iteratively moved towards the learned data manifold using a deterministic score-based denoising term and corrupted again by a stochastic noise injection term. The latter helps correct errors made in earlier sampling steps and prevents solutions from solely converging to high-density regions \cite{songGenerativeModelingEstimating2019, karrasElucidatingDesignSpace2022}.

\subsection{Ultrasound Dehazing}
\label{sec:dehazing}
In this section, we proceed with technical details on how we use diffusion models to perform posterior sampling for the dehazing task. We aim to condition the sampling process on a given measurement $\by$, i.e. the hazy acquisition. Additionally, we provide practical advancements that were necessary to efficiently learn and apply priors for ultrasound data. 
\subsubsection{Joint Posterior Sampling}
Two prior distributions can be combined during posterior sampling by sampling from them in parallel whilst conditioning on the measurement, which was shown in \cite{stevensRemovingStructuredNoise2023}. By incorporating an explicit noise prior, we are able to learn and utilize complex noise distributions $p_H(\bh)$ to effectively eliminate any such noise from the measurement $\by$. In terms of diffusion modeling, this joint posterior sampling process $p_{X|Y}(\bx, \bh|\by)$ is achieved through the formulation of a \emph{joint conditional} diffusion process $\left\{\bx_t, \bh_t|\by\right\}_{t\in[0,1]}$, in turn producing a joint conditional \emph{reverse-time} SDE:
\begin{align}
    \mathrm{d}(\bx_t, \bh_t) &= 
    \big[
        f(t)(\bx_t, \bh_t) - \ldots \nonumber\\ &g(t)^2 
            \nabla_{\bx_t, \bh_t} \log{p(\bx_t, \bh_t|\by})
    \big] \mathrm{d}t + g(t) \mathrm{d}\Bar{\mathbf{w}_t},
    \label{eq:cond_reverse_diff}
\end{align}
which is essentially an extension of \ref{eq:reverse_diff}, with the inclusion of the haze distribution (joint) and conditioning on the observation (conditional). Instead of the prior gradient that was derived in the unconditional sampling case, see \eqref{eq:reverse_diff}, the posterior gradient appears. We can apply the Bayesian factorization in \eqref{eq:bayes} to construct two separate diffusion processes, defined by separate score models but entangled through our haze forward model \TSWR{$p_{Y|X,H}(\by|\bx, \bh)$}, see \eqref{eq:inverse_model}. In addition to the original score model $s_\theta(\bx, t)$, we introduce a second score model $s_\phi(\bh_t, t)\simeq \nabla_{\bh_t} \log p_H(\bh_t)$, parameterized by weights $\phi$, to model the haze. These two score networks can be trained on clean ultrasound and haze datasets independently, using the objective in \eqref{eq:score_matching}. The gradients of the posterior with respect to $\bx$ and $\bh$ are now given by:
\begin{equation}
\label{eq:approx_cond}
    \left\{
        \begin{array}{lr}
                \nabla_{\bx_t}\log{p(\bx_t, \bh_t|\by)} \simeq s_\theta^*(\bx_t, t) + \nabla_{\bx_t} \log{p(\by|\bx_t, \bh_t)}
                \\[4pt]
                \nabla_{\bh_t}\log{p(\bx_t, \bh_t|\by)} \simeq s_\phi^*(\bh_t, t) + \nabla_{\bh_t} \log{p(\by|\bx_t, \bh_t)}.
        \end{array}
    \right.
\end{equation}
Substituting the posterior in our reverse-time SDE\\ formulation allows us to sample from the posterior and effectively obtain the dehazed output $\bx$ as well as an estimation of the haze $\bh$. It should be noted that the true noise-perturbed likelihood $p(\by|\bx_t, \bh_t)$, which appears in \eqref{eq:approx_cond}, is generally intractable, unlike $p(\by|\bx_0, \bh_0)$. Different approximations have been proposed by \cite{chungDiffusionPosteriorSampling2023}, \cite{mengDiffusionModelBased2023} and \cite{fengScoreBasedDiffusionModels2023}. In this work, we opt for a solution used in \cite{songScoreBasedGenerativeModeling2021}, which essentially corrupts the observation $\by$ along the diffusion process $\left\{\by_t\right\}_{t\in[0,1]}$ to obtain the projected value $\hat{\by}_t\sim q(\by_t|\by_0)$ which results in an approximated version of the noise-perturbed likelihood \TSW{$p(\by|\bx_t, \bh_t) \approx p(\hat{\by}_t|\bx_t, \bh_t) = \mathcal{N}(\bx_t + \bh_t, \rho^2 \bI)$, which we assume to be Gaussian with diagonal variance $\rho^2$. Although more sophisticated likelihood approximations for the joint diffusion method have been explored in \cite{stevensRemovingStructuredNoise2023}, we find that the aforementioned method is sufficient for the dehazing task in this study, while also being computationally less expensive.}

\subsubsection{Learning Ultrasound Priors}
\begin{figure}
    \centering \includegraphics[clip, trim=0 0.9cm 0 0.4cm]{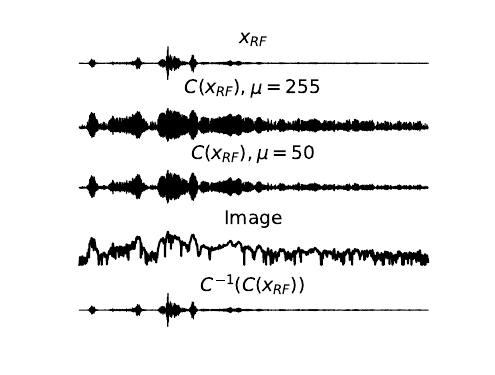}
    \caption{RF signals alongside their companded versions. The $\mu$ value is tuned to more closely resemble the distribution of the image pixel values.}
    \label{fig:companded-signals}
    \vspace{-0.5cm}
\end{figure}

\begin{figure}
    \centering
    \includegraphics[width=\columnwidth]{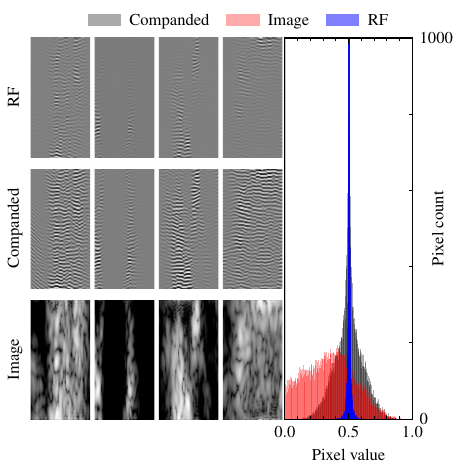}
    \caption{A comparison of radio-frequency (RF), companded RF, and image of four arbitrary ultrasound patches from the \emph{in-vivo} training dataset (left to right). A histogram comparison reveals the high dynamic range of the RF data which after normalization predominantly occupies the center region of pixel values.}
    \label{fig:histogram}
    \vspace{-0.5cm}
\end{figure}

The vast majority of applications that involve learning generative models are trained in the image domain. Subsequently, the design choices that work well with image data are well-established. However, when it comes to ultrasound data, it is less obvious how to present the data. Learning priors on pre-envelope detected ultrasound data requires careful design of the preprocessing stages and generative network. In practice, we found that the straightforward adoption of current methods for image data did not translate well to the ultrasound domain. Most notably, the high dynamic range of ultrasound data is an issue for training a prior. Fig.~\ref{fig:histogram} shows a histogram comparison between raw RF and image ultrasound data. Because the majority of the RF samples are concentrated around the center pixel value, activation functions within the neural network are not operated in their optimal range, which stagnates the training progress. This makes it impractical to learn generative models on raw ultrasound data. To that end, we borrow a technique named companding, that originates from the telecommunication and audio fields \cite{sklar2021digital}. Companding is an invertible operation that can \emph{compress} and \emph{expand} the dynamic range of a signal. We here focus on the $\mu$-law companding algorithm, consisting out of the compression, i.e. $\mu$-law encoding:
\begin{align}
    C(\bx_{\text{RF}}) = \sign(\bx_{\text{RF}})\frac{\ln({1 + \mu \vert \bx_{\text{RF}} \vert)}}{\ln({1 + \mu)}}, \quad -1\leq \bx_{\text{RF}} \leq 1,
\end{align}
and its inverse, the $\mu$-law expansion being:
\begin{align}
    C^{-1}(\bx) = \sign(\bx)\frac{(1 + \mu)^{\vert \bx \vert} - 1}{\mu}, \quad -1\leq \bx \leq 1,
\end{align}
where $\mu=255$ is a parameter that determines the amount of compression applied, see Fig~\ref{fig:companded-signals}, and $\bx$ should be normalized first to the range $[-1, 1]$. Effectively, companding will transform the signal to the logarithmic domain (positive and negative parts piecewise using the sign function $\sign(\cdot)$), reducing its dynamic range. It has the useful property of invertibility such that $\bx_\text{RF} = C^{-1}(C(\bx_\text{RF}))$. An example of the companded data is shown in Fig.~\ref{fig:histogram}, which resembles the distribution of image pixel values more closely compared to plain RF data.

Most conditional or guided diffusion methods focus on problems where the conditioning is applied in the same domain or a linear transformation thereof as the learned prior. However, in the case of ultrasound dehazing, as formulated in this work, the measurement domain (RF) is different from the domain of the diffusion trajectory (companded RF). We therefore expand the companded RF data during the data consistency step to ensure it conforms with \eqref{eq:inverse_model}
\begin{align}
    \TSW{\nabla_{\bx_t}\log} & \TSW{\, p(\by|\bx_t, \bh_t) \approx \nabla_{\bx_t}\log \, p(\hat{\by}_t|\bx_t, \bh_t) = \ldots} \nonumber\\
    &= \lambda \nabla_{\bx_t} \norm{\hat{\by}_t - C( \bx_{\text{RF}, t} + \bh_{\text{RF}, t})}_2^2 \nonumber\\
    &= \lambda \nabla_{\bx_t} \norm{\hat{\by}_t - C(C^{-1}(\bx_t) + \gamma C^{-1}(\bh_t))}_2^2,
\end{align}
\TSW{where $\lambda$ encompasses the likelihood variance $\rho^2$ and can also be seen as a weighting term for the importance of the measurement error with respect to the prior, which is common practice in Bayesian inference.} Furthermore, we introduce a parameter $\gamma$ which effectively specifies the desired signal-to-haze ratio. Notice how the $\ell_2$-norm is applied in the companded domain, rather than the RF domain, to align the gradients with the score steps. A similar approach is followed to arrive at the likelihood term for the haze component $\nabla_{\bh_t}\log p(\hat{\by}_t|\bx_t, \bh_t)$, alongside the introduction of a second weighting term $\kappa$. We use Tensorflow's automatic differentiation to automatically compute the gradients for the data consistency steps. See Algorithm~\ref{alg:joint_cond_sampler} for a more complete overview of the dehazing by joint posterior sampling method.
\begin{algorithm}[tb]
    \caption{Dehazing by joint posterior diffusion sampling}
    \label{alg:joint_cond_sampler}
\begin{algorithmic}[1]
    \REQUIRE $T, s_\theta, s_\phi, \lambda, \kappa, \by, t_\tau, \gamma$
    \SET $\Delta t \gets \frac{1}{T}, \bz \sim \mathcal{N}\left(0, \bI\right)$
    \SET $\by = C(\by_\text{RF})$ \algocomment{companding}
    \SET $[ \by^{(0, 0)}, \ldots, \by^{(n, m)}, \dots, \by^{(N, M)} ] \gets \by$ \algocomment{factorize}
    \SET $\bx_{t_\tau} \gets \alpha_{t_\tau}\by + \beta_{t_\tau} \bz, \quad \bh_{t_\tau} \gets \bx_{t_\tau}$
    \algocomment{initialization}\\
    \algocomment{Reverse diffusion steps}
    \FOR{$t=t_\tau$ {\bfseries to} $0$ {\bfseries step} $\Delta t$}
    
        \STATE \algocomment{Loop through patches}
        
        \FOR{$n=0, m=0$ {\bfseries to} $N, M$}
        
        
            \STATE \algocomment{Denoting $\bx^{(m, n)}$ as $\bx$ for shorthand}
            \STATE $\log{p(\hat{\by}_t|\bx_t, \bh_t)} \gets \ldots \newline
            \hspace*{5mm} \norm{\hat{\by}_t - C(C^{-1}(\bx_t) + \gamma C^{-1}(\bh_t))}_2^2$
            
            \STATE \algocomment{Data consistency steps}
            \STATE $\bx_t \gets \bx_t + \lambda \nabla_{\bx_t} \log{p(\hat{\by}_t|\bx_t, \bh_t)}$
            \STATE $\bh_t \gets \bh_t + \kappa \nabla_{\bh_t} \log{p(\hat{\by}_t|\bx_t, \bh_t)}$
            
            \STATE \algocomment{Reverse diffusion steps}
            
            \STATE $\bx_{t-\Delta t} \gets \bx_{t} - f(t)\bx_{t} \Delta t$ 
            
            \STATE $\bx_{t-\Delta t} \gets \bx_{t-\Delta t} + g(t)^2 s_\theta^*(\bx_{t}, t) \Delta t$\\
            \STATE $\mathbf{z} \sim \mathcal{N}(\mathbf{0}, \mathbf{I})$\\
            \STATE $\bx_{t-\Delta t} \gets \bx_{t-\Delta t} + g(t) \sqrt{\Delta t} \mathbf{z}$
            
            \STATE \;
            \STATE $\bh_{t-\Delta t} \gets \bh_{t} - f(t)\bh_{t} \Delta t$
            
            \STATE $\bh_{t-\Delta t} \gets \bh_{t-\Delta t} + g(t)^2 s_\phi^*(\bh_{t}, t) \Delta t$\\
            \STATE $\mathbf{z} \sim \mathcal{N}(\mathbf{0}, \mathbf{I})$\\
            \STATE $\bh_{t-\Delta t} \gets \bh_{t-\Delta t} + g(t) \sqrt{\Delta t} \mathbf{z}$ 
         
            \STATE \algocomment{Dropping shorthand notation}
            \STATE \algocomment{Interleave patches}
            \FOR{A {\bfseries in} \small{$\left\{(n, m{-}1), (n{-}1, m), (n{-}1, m{-}1)\right\}$}}
                \STATE $\bx_{t}^{A\,\cap\,(n, m)} =  \bx_{t}^{(n, m)\,\cap\,A}$
                \STATE $\bh_{t}^{A\,\cap\,(n, m)} =  \bh_{t}^{(n, m)\,\cap\,A}$
            \ENDFOR 
        \ENDFOR 
    \ENDFOR 
    \STATE $\bx_0 \gets [ \bx^{(0, 0)}, \dots, \bx^{(N, M)} ]$ \algocomment{patch stitching}
    \STATE $\bx_\text{RF} \gets C^{-1}(\bx_0)$ \algocomment{expanding}
    \OUTPUT $\bx_\text{RF}$
\end{algorithmic}
\end{algorithm}

\subsubsection{Patch Based Inference}
\label{sec:patch_based}
The prior distribution can be factorized into multiple patches for computational efficiency (reducing memory footprint) and better generalization of the generative network. \TSW{Regarding the latter, a patch-based approach can mitigate overfitting to specific anatomical structures and geometries in the train set. The patch-based inference effectively reduces the receptive field of the network, which has been shown to mitigate overfitting to the train set and improve generalization \cite{koutiniReceptiveFieldRegularization2021}}.

\begin{figure}
  \centering
   \includegraphics[width=0.48\columnwidth]{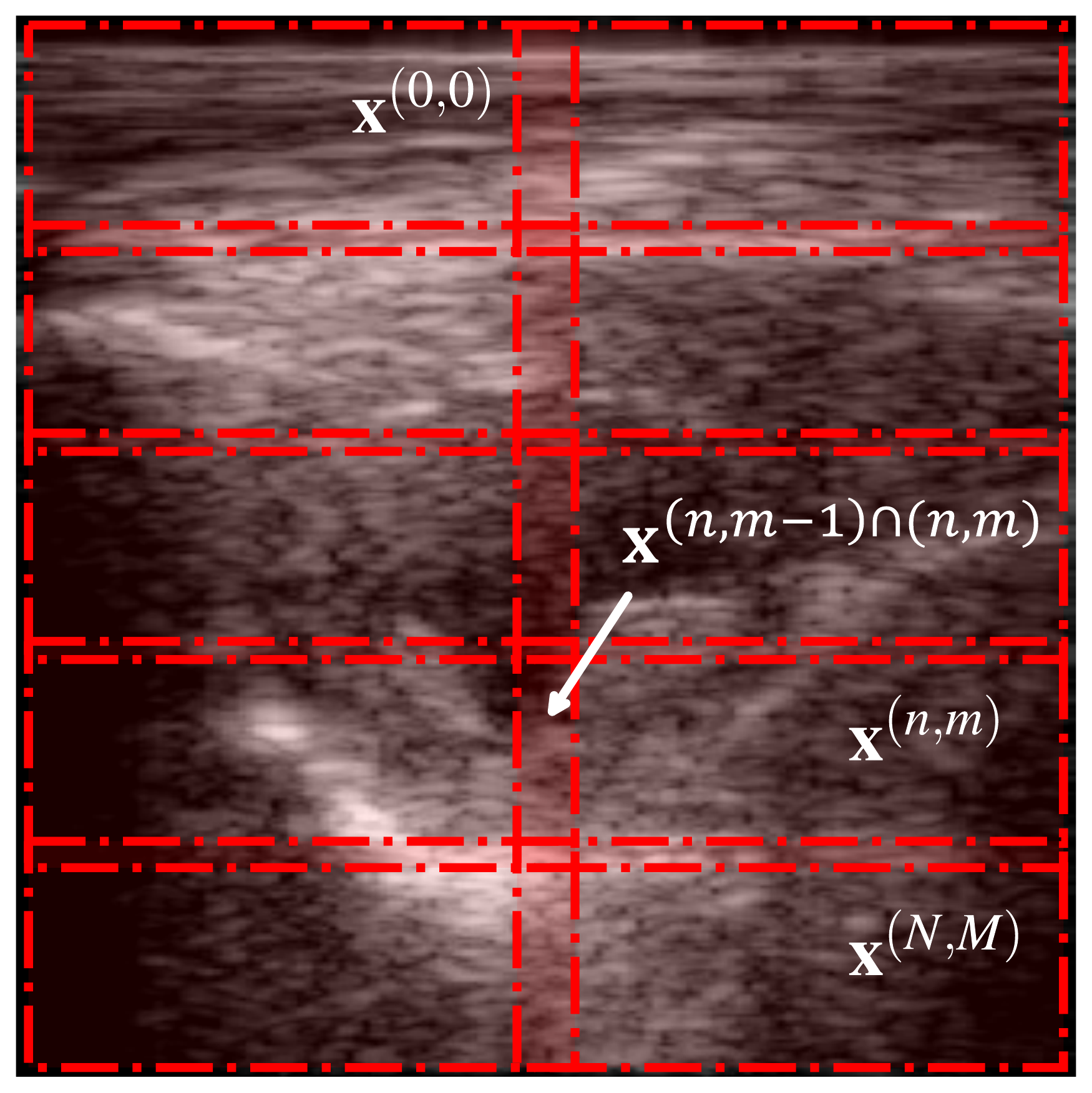}
  \caption{Patch-based inference with $N\times M$ overlapping patches.}
  \label{fig:patch_inf}
  \vspace{-0.5cm}
\end{figure}

Independently running inference on each patch separately seems a straightforward solution, but results in unnatural stitch artifacts - even when overlapping patches and window-based averaging are applied. Diffusion models allow for more sophisticated schemes to generate larger content \cite{zhangDiffCollageParallelGeneration2023}. A simple yet effective solution is the mask-shift trick, introduced by \cite{wangZeroShotImageRestoration2022}, which solves this issue by enforcing coherence between reconstructed patches. It interleaves the diffusion process of neighboring patches, by replacing overlapping parts of the current patch with the adjacent patches at each time step of the iterative algorithm. Inspired by this, we apply a similar technique by factorizing the observation $\by$ into $N\times M$ patches with some overlap between adjacent patches, see Fig.~\ref{fig:patch_inf}:
\begin{align}
    \by = [ \by^{(0, 0)}, \ldots, \by^{(n, m)}, \dots, \by^{(N, M)} ].
\end{align}
We perform a single conditional diffusion step on all patches separately which results in samples $[ \bx_{t}^{(0, 0)}, \ldots,\bx_t^{(N, M)} ]$. Next, we replace all overlapping pixels of the adjacent patch with the current patch as follows:
\begin{align}
    \bx_{t}^{(n, m-1)\cap(n, m)} =  \bx_{t}^{(n, m) \cap (n, m-1)},
    \label{eq:patch_replacement}
\end{align}

where we use the intersection symbol $A\cap B$ to denote the pixels from patch $A$ that overlap with pixels from patch $B$. In this case, \eqref{eq:patch_replacement} shows the patch replacement for the patch to the left of the current patch $(n, m-1)$. Naturally, we also apply the patch replacement to the other adjacent patches at indices $(n-1, m-1)$ and $ (n-1, m)$. After $T$ diffusion steps, we can reconstruct the dehazed image by combining the overlapping patches as follows:
\begin{align}
    \bx_0 = [ \bx^{(0, 0)}, \dots, \bx^{(N, M)} ].
\end{align}
Because of the interleaving of the patches with the diffusion process, no windowing or averaging has to be applied to the overlapping parts, as they are already consistent. Note that we can dehaze the individual patches in parallel and are not reliant on autoregressive methods which would substantially slow down the inference \cite{sahariaPaletteImagetoImageDiffusion2022}.

\subsubsection{Haze Estimation}
\label{sec:haze_estimation}
The proposed inference scheme considers the haze to be unknown, and it is not necessary to compute any local noise statistics to do haze estimation at test time as we will simply use the learned haze prior. Using generative modeling, the full complexity of the haze distribution can be captured and applied, without relying on basic assumptions or estimations during inference. However, learning a prior on the haze signal requires a collection of (surrogate) haze signals $\mathcal{H} = \left\{\bh_0^{(1)}, \bh_0^{(2)}, \ldots,  \bh_0^{(|\mathcal{H}|)}\right\}\sim p(\bh_0)$ to learn the haze distribution in advance (train time). The advantage is that we do not require ultrasound measurements with matching haze signals. A collection of stand-alone haze signals will suffice, which provides great flexibility in how the haze dataset can be acquired. 

For instance, hazy signals can be independently acquired in an experimental environment using a dedicated haze phantom, such as a stainless steel scouring pad in an empty water tank \cite{vignonRevisitingWienerPostfilter2020}. The phantom induces high-order multiple scattering, simulating the multipath reflections within the chest wall, which are the major contributors to the haze signal \cite{fatemiStudyingOriginReverberation2019}.

Alternatively, haze signals can be estimated from a noisy measurement by extracting off-axis energy, mainly side lobes and clutter. Using two inverted apodization schemes, the main lobes can be canceled by coherent subtraction of the two separately beamformed images. Assuming haze scatters are detected from every direction, in low SNR regions the off-axis energy is dominated by clutter and can be used to estimate the haze. To further ensure minimal signal leakage into the haze dataset, we leverage the fact we are training on patches and preprocess the dataset by sampling patches that are more likely to contain haze using focus measures, as was done in \cite{vandeschaftUltrasoundSpeckleSuppression2021}.

\subsubsection{Initialization}
As presented in \cite{chungComeCloserDiffuseFasterAcceleratingConditional2022}, come closer diffuse faster (CCDF) provides a superior initialization method compared to sampling from the base distribution $\bx_1\sim\pi(\bx)$. CCDF is an accelerated sampling scheme that starts the reverse diffusion at some time $ t_\tau$ where $t_0 \leq t_\tau \leq t_T$, which simultaneously reduces the number of diffusion steps necessary to $t_\tau \cdot T$ and provides a better initial estimate derived from the measurement through forward diffusion $\bx_{t_\tau} = \alpha_{t_\tau}\by + \beta_{t_\tau} \bz$. We here extend CCDF to work with the proposed joint posterior sampling scheme. Both clean ultrasound signal $\bx_{t_\tau}$ and haze $\bh_{t_\tau}$ initial estimates are initialized using a forward diffusion step towards $t=\tau$ from the measurement $\by$. \TSW{This initialization allows us to reduce the number of sampling steps, hence speeding up the inference procedure. $\tau$ is chosen in such a way that no noticeable loss in reconstruction quality is observed in comparison with Gaussian initialization.}

\subsubsection{Training and Inference Details}
\label{sec:training_details}
For both the tissue and haze score models, we use the NCSNv2 architecture as introduced in \cite{songImprovedTechniquesTraining2020}. Both score models are trained on patch datasets of size $128\times64$ denoting axial and lateral dimensions, respectively \TSW{for 100 epochs, with batch size of 8 images and learning rate of \num{1e-4}.} During training, augmentation is applied through a random left-right flip with equal probabilities and a random brightness offset uniformly sampled between the values $\pm 0.1$ given an image range of $\bx\in[0, 1]$. The dataset is normalized to this range based on its extreme values. In the rare event the augmentation exceeds the set normalized range, values are clipped. During inference, the two trained networks are combined through the proposed sampling procedure as described in Algorithm~\ref{alg:joint_cond_sampler}. We use the following SDE: $f(t)=0$, $g(t)=\sigma ^ t$ with $\sigma=25$ to define the diffusion trajectory, resulting in $\alpha_t = 1$ and $\beta_t = \frac{1}{2\log\sigma}(\sigma^{2t}-1)$. \TSWR{During each experiment, we run the sampler for $T=200$ steps, starting at $\tau=0.8$, which proved to be sufficiently long for producing adequate samples.} Empirically, we found that weighting values close to $\lambda\approx0.5$ and $\kappa\approx0.5$ work well. We use a 10\% overlap between adjacent patches to interleave the diffusion process as described in Section~\ref{sec:patch_based}. To emphasize a fair comparison, we perform hyperparameter tuning for all experiments and (baseline) methods only on a small subset of 5 images. \TSW{All experiments and training of the models were done on a single 12GBytes NVIDIA GeForce RTX 3080 Ti. The implementation was carried out using \TSWR{Python 3.10 and TensorFlow 2.10}. The code for the joint diffusion algorithm is made publicly available\footnote{dehazing-diffusion.github.io}.}

\section{Experiments}
\label{sec:results}
To test the proposed method, we conducted two different experiments. In the first experiment, we set up a controlled environment with heart and haze phantoms (\emph{in-vitro}). In the second experiment, we use hazy \emph{in-vivo} cardiac ultrasound data. All experimental data is recorded using an X5-1C matrix phased array transducer connected to a Philips EPIQ scanner (Philips Research, North America). The transducer has a frequency range of \qtyrange{1}{5}{\mega\hertz} and all acquisitions are made using harmonic imaging.
 
Even though the proposed dehazing method operates in the RF domain, all metrics (see sections \ref{sec:invitro}, \ref{sec:invivo}) are computed after log-compression and envelope-detection (B-mode), since the images are eventually displayed in that format. Lastly, all images are brightness matched by matching the average intensity of the top 10\% pixel values and plotted in the same dynamic range of \qty{60}{dB} for a fair visual comparison. 

\begin{figure*}
    \centering
    \includegraphics[width=\textwidth, clip, trim=0 0.3cm 0 0]{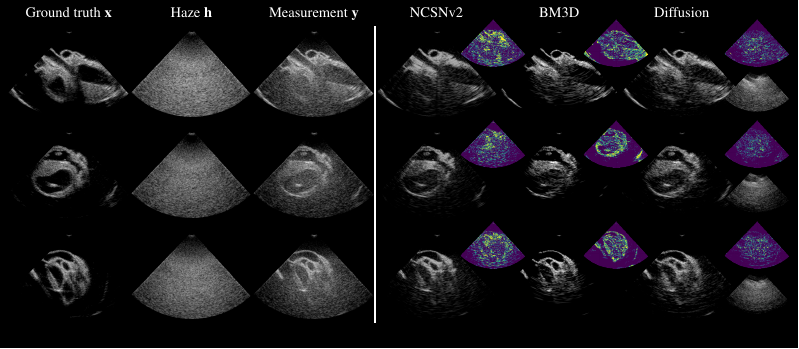}
    \caption{\emph{In-vitro} results using the phantom data. The ground truth ultrasound $\bx$ and haze signals $\bh$ which are used to construct measurement $\by$ are shown on the left. RF-based dehazing results are on the right, for both the baseline methods (BM3D and NCSNv2) and the proposed diffusion method. For the latter, we show posterior solutions for both the ultrasound and haze signals (lower inset plot to the right). For all methods, we show error plots that highlight the difference between ground truth and dehazed images, with the proposed diffusion method notably dropping less signal.}
    \label{fig:comparison-invitro}    
\end{figure*}
\subsection{In-Vitro}
\label{sec:invitro}
To thoroughly examine the method's behavior, we conducted a controlled test using a heart phantom and a dedicated haze phantom (as described in Section~\ref{sec:haze_estimation}). This allowed us to collect two distinct sets of beamsummed RF data. These datasets were utilized to train the two diffusion models, $s_\theta$ and $s_\phi$, respectively. We acquired a total of $2\times150$ frames for the training dataset and $2\times38$ frames for validation, of both the heart and haze phantoms separately. During inference, we artificially add the haze signal to the heart phantom ultrasound data with \TSW{varying haze levels} to imitate the haze process and set the desired haze-to-signal ratio. This enables us to compare the dehazed output to the ground truth, ensuring that the extracted haze aligns with the desired haze distribution.
\vspace{-0.1cm}

\subsection{In-Vivo}
\label{sec:invivo}
In a second experiment, we train and deploy our method on an \emph{in-vivo} cardiac ultrasound dataset. In total, we acquired ultrasound data from five different human volunteers and used four different commonly used views to record the echocardiograms. Namely the apical three- and four-chamber view (AP3/AP4), parasternal long axis (PLAX), and short axis (PSAX) views. \TSW{Combined, the dataset comprises a total of 2640 frames collected from 6 volunteers. All 44 acquisitions consist of 60-frame sequences, each covering approximately a complete cardiac cycle. We divided this dataset into three main subsets: 1) a training set, earlier denoted as $\mathcal{X}$, consisting of 1500 clean frames from three volunteers which are used for training the tissue model, 2) a validation set comprising of 1020 frames from two other volunteers for validation of the proposed method, 3) another training set with the remaining 120 frames of the last volunteer. The latter two datasets are considered hazy and were acquired from technically difficult-to-image subjects. To generate the specialized haze-only training dataset $\mathcal{H}$ for training the haze model, we use the acquisitions from the third dataset combined with the relatively hazy parts of the first training dataset for augmentation. Subsequently, the generation of the haze-only images involves the apodization method outlined in Section~\ref{sec:haze_estimation}. This method extracts off-axis clutter by cancelation of the main lobes, effectively eliminating tissue signal and concentrating solely on haze characteristics. Lastly, as part of our methodology, we deliberately reserved 5 images from each of the tissue (1) and haze (3) training datasets. These images are exclusively used for fine-tuning both training and inference hyperparameters.}

\subsection{Metrics}
As a quantitative measure, we compute the peak signal-to-noise ratio (PSNR) between the dehazed output and ground truth images. The PSNR is a pixel-wise difference metric that normalizes the mean squared error between two images by the image range to account for different intensity ranges across examples. Unlike the \emph{in-vitro} experiment, there is no ground truth available for the \emph{in-vivo} data. Therefore, we compute the generalized contrast-to-noise ratio (gCNR) \cite{rodriguez-molaresGeneralizedContrasttonoiseRatio2020}, which is an unsupervised image quality metric, commonly used in ultrasound, given by:

\begin{equation}
    \mathrm{gCNR}(I) = 1 - \int_x \min\left\{ p_A(x), p_B(x) \right\} \mathrm{d}x,
\end{equation}
where $p_A$ and $p_B$ are the probability density functions for each of the regions of interest (ROIs). We compute a histogram for each ROI to estimate its distribution and choose the heart chamber (ventricle) near the apex (typically the most challenging and most hazy region) as region $A$ and part of the ventricular septum (wall) as region $B$. Most notably, the gCNR is not biased by dynamic range stretching and can thus provide a fair comparison independent of specific brightness settings. 

\TSW{To ensure accurate results, we manually draw two masks as ROIs for seven equidistant frames in each of the 60-frame sequences in the dataset, using a lasso selection tool. For the remaining frames, the masks are determined by linearly interpolating the vertices of a polygon, which are fitted to each mask using the Douglas-Peucker algorithm. This approach provides us with larger and more precise regions that move in} \TSW{synchronization with the cardiac cycle, as opposed to relying on a single set of ROIs per sequence. On average, the areas for regions $A$ and $B$ are \qty{14}{\cm\tothe{2}} and  \qty{8}{\cm\tothe{2}}, respectively}. We remove outliers by eliminating the scores below the $10$th percentile obtained from the measurement frames, as these frames are assumed to have inaccurate ROIs. Since we remove outliers based on scores from the raw harmonic data, this should not favor any of the proposed or baseline methods.

\TSW{Additionally, we utilize the Kolmogorov-Smirnov (KS) test to assess the ability of dehazing methods to retain speckle statistics in the myocardium regions. Furthermore, we investigate how the dehazing methods affect the ventricle region, which is expected to predominantly contain haze and no tissue signal. The KS statistic is defined as the greatest distance between the empirical distributions of two samples and serves as a measure of statistical similarity.} 

\TSW{Lastly, since the KS test only considers pixel statistics independently, we more closely investigate speckle characteristics, by estimating the speckle size. This provides us with a measure of the lateral resolution and insight into the spatial} \TSW{preservation of statistics. The speckle size is evaluated through measurement of the Full-Width at Half-Maxima (FWHM) of the lateral main lobe of the two-dimensional autocorrelation taken from the myocardium region in the polar domain. The same ROIs as for the gCNR computation are used for both the KS test and the FWHM estimation.}\color{black}

\begin{figure}
    \centering
    \includegraphics[scale=0.145, trim=2cm 3cm 0 0]{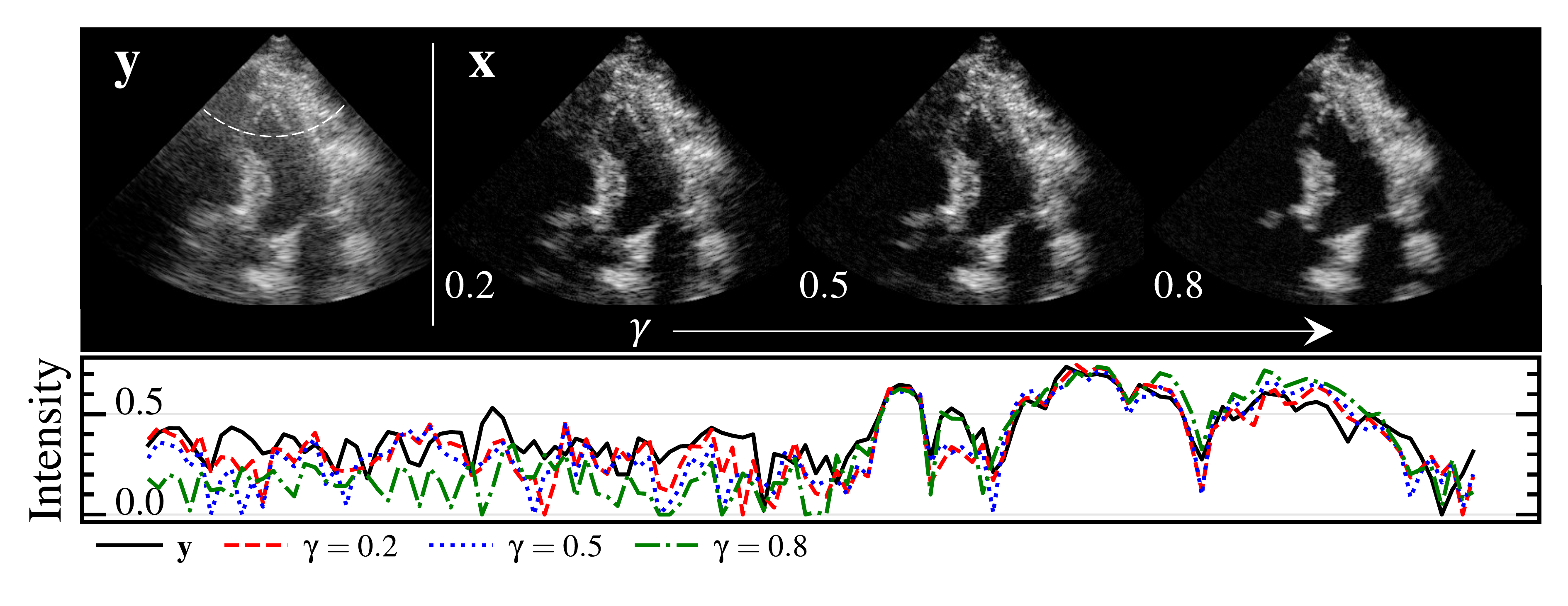}
    \caption{Varying amount of dehazing using tunable haze amplitude parameter $\gamma$. A lateral cross-section is shown for a more detailed comparison. The exact slice is indicated by the white dashed line.}
    \label{fig:varying_gamma}
\end{figure}
\subsection{Baselines}
As a baseline method, we use the block-matching and 3D filtering algorithm (BM3D) \cite{dabovImageDenoisingBlockmatching2006}. BM3D has been a popular noise suppression algorithm and has recently been adopted for ultrasound \cite{santosUltrasoundImageDespeckling2017}. BM3D outperforms transform-based denoisers, which often fail to preserve details that are not represented in the 2D transform domain. Furthermore, BM3D is general concerning the type of noise. We adapted and fine-tuned BM3D for the cardiac dehazing problem to create a competitive baseline. To that end, we deploy BM3D in the RF domain, as the image domain results mainly introduce blurring of the images without any dehazing effect.

Additionally, we compare the proposed technique with a deep learning method. Specifically, we use a supervised deep learning approach, training the model end-to-end to tackle the dehazing task on a paired dataset. Although not readily available in the literature for dehazing, the U-Net has been the go-to neural architecture in medical image restoration. For a fair comparison, we use the same backbone network, NCSNv2, used in our diffusion approach, which is also a U-Net type network. We train the supervised network on the RF phantom dataset, as we only have pairs for this dataset. This is the inherently limiting part of the supervised approach.

\subsection{Downstream Delineation Task}
\label{sec:downstream}
\TSW{In the context of ultrasound imaging for cardiac assessment, downstream tasks can serve as valuable metrics to quantify the effectiveness of dehazing algorithms presented in this paper. Besides automated tasks, manual labeling of cardiac images can be performed easier and quicker given improved image quality. One common cardiac measurement involves the delineation of the ventricle, which is used for the calculation of parameters like ejection fraction \cite{moal2022explicit}. Ejection fraction is a key indicator of heart function and is used to assess the heart's ability to pump blood efficiently. Accurate assessment of the volume of the cardiac chamber is critical for proper diagnoses. In this experiment, we use a pretrained EchoNet~\cite{ouyangVideobasedAIBeattobeat2020} for delineation of the left ventricle and compare masks produced using the unprocessed harmonic imaging data as input with masks produced using diffusion-processed images. For this experiment, we filter our dataset on apical views only as EchoNet was trained on these. Furthermore, we slightly changed the masks used in the other experiments to better match the masks produced by EchoNet, which often included the valves (which we want to exclude for proper gCNR estimation).}

\begin{figure}
    \centering
    \includegraphics[trim=0 0.5em 0 0]{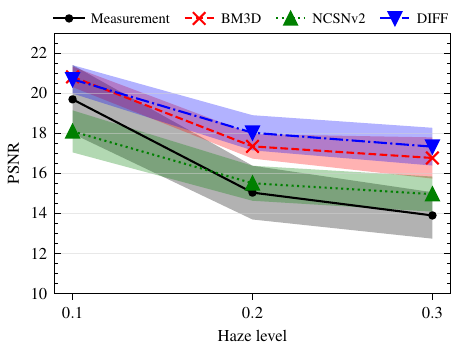}
    \caption{Comparison of dehazing methods with harmonic imaging for varying levels of haze in the \emph{in-vitro} experiment. Shown are the average and standard deviation (shaded areas) of the PSNR over all frames. All pairs of PSNR scores pass the Welch’s t-Test for statistical significance ($p<0.05$), except for BM3D and DIFF at haze level $0.1$, and NCSNv2 and measurement at haze level $0.2$}
    \label{fig:haze_levels}
    \vspace{-0.5cm}
\end{figure}

\section{Results}
\color{black}

\subsection{In-Vitro}
\label{sec:res-invitro}
Fig.~\ref{fig:comparison-invitro} shows dehazing results on the phantom data. The RF-based diffusion dehazing method can reconstruct realistic dehazed images with improved contrast as well as provide plausible haze estimations. Both baseline methods (BM3D, NCSNv2) can remove most of the haze but at the cost of dropping signal from low-level tissue (typically endocardium). \color{black} This is often not acceptable in clinical practice as it hinders proper contouring of the left ventricle, in turn prohibiting accurate volume analysis or tracking for strain estimation. Lastly, we compare the proposed method against the baseline at various haze levels in Fig.~\ref{fig:haze_levels} and see an improved PSNR score across all haze levels.
\subsection{In-Vivo}
\begin{figure}
    \centering
    \includegraphics[trim=0 0 0 0.1cm]{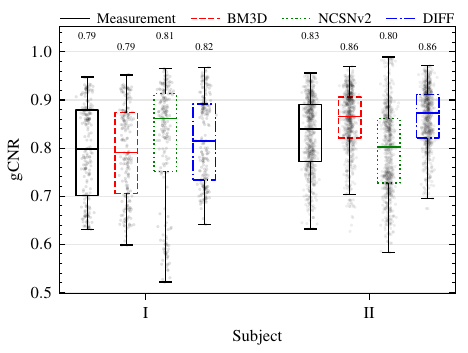}
    \caption{gCNR scores on the \emph{in-vivo} dataset grouped into separate subjects, comparing the raw harmonic images with baseline and proposed methods. Each dot represents a time frame within the acquisitions.}
    \label{fig:gcnr_subject}
\end{figure}

\begin{figure}
    \centering
    \includegraphics[trim=0 0 0 0.1cm]{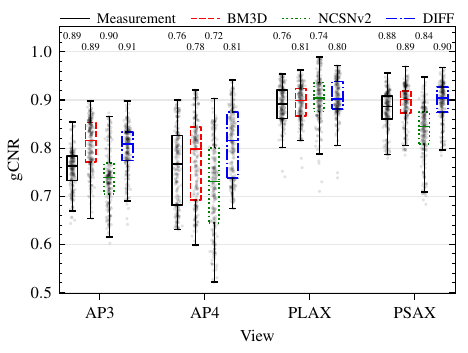}
    \caption{gCNR scores on the \emph{in-vivo} dataset grouped into the four different views, comparing the raw harmonic images with baseline and proposed methods. Each dot represents a time frame within the acquisitions.}
    \label{fig:gcnr_view}
    \vspace{-0.5cm}
\end{figure}

\begin{figure}
    \centering
    \includegraphics[trim=0 0.1cm 0 0]{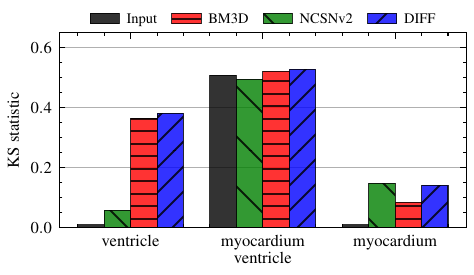}
    \caption{\TSW{Kolmogorov-Smirnov statistic for testing statistical similarity of the speckle in the myocardium and ventricle regions, pre- and post-dehazing. From left to right, we compare each of the two different regions of dehazing methods with the original data. The diffusion method retains the speckle statistics in the myocardium region, while clearly altering the contents from the ventricle. In contrast, the NCSNv2 method is not able to suppress haze in the ventricle area, leading to more similar statistics with respect to the input. As a reference, we also compare KS statistics between the dehazed ventricle and myocardium regions (middle column), which naturally are statistically different.}}
    \label{fig:ks-test}
\end{figure}

\begin{figure}
    \centering
    \includegraphics{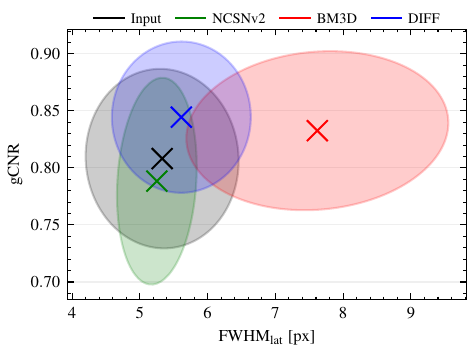}
    \caption{\TSW{Depiction of the FWHM of the myocardium speckle against the gCNR scores, depicting the inherent trade-off between contrast and lateral resolution. Each ellipse represents the average and standard deviation of each metric for that particular model. Our analysis reveals that the diffusion method both obtains high contrast and preserves lateral resolution, while the BM3D method compromises significantly on resolution.}}
    \label{fig:fwhm_lat_gcnr}
\end{figure}

\color{black}
\label{sec:res-invivo}
The quantitative results on the dehazing of the \emph{in-vivo} are summarized in Fig.~\ref{fig:gcnr_subject} and \ref{fig:gcnr_view}, respectively. Fig.~\ref{fig:gcnr_subject} shows gCNR scores for both of the difficult-to-image subjects in the validation set, while Fig.~\ref{fig:gcnr_view} shows results for each of the four different acoustic windows. The spread in scores for each method can be mainly attributed to the variety of haze severity across sequences. Nonetheless, a clear pattern in the gCNR scores is detected. Compared to straightforward harmonic imaging, the BM3D method results in comparable and sometimes lower gCNR scores. This can be attributed to the excessive removal of signal by BM3D, which was also seen in the \emph{in-vitro} experiment. In contrast, the proposed diffusion method yields improved gCNR scores. Unlike the relatively good performance of the NCSNv2 supervised method on the phantom data, it fails to effectively remove the haze on the \emph{in-vivo} data, as expected for out-of-distribution data. Fig.~\ref{fig:ks-test} shows the KS test results on the \emph{in-vivo} dataset. Most notably, it can be seen that the proposed method retains speckle statistics in the myocardium region while altering the contents of the ventricle (read dehazing). Concerning the lateral resolution of the dehazing methods, the diffusion method strikes a balance between improving contrast, without sacrificing spatial resolution, see Fig.~\ref{fig:fwhm_lat_gcnr}. In comparison, the BM3D method compromises significantly on resolution, which is a common problem in denoising algorithms due to their smoothening nature.

\begin{figure*}
    \centering
    \includegraphics[width=\textwidth]{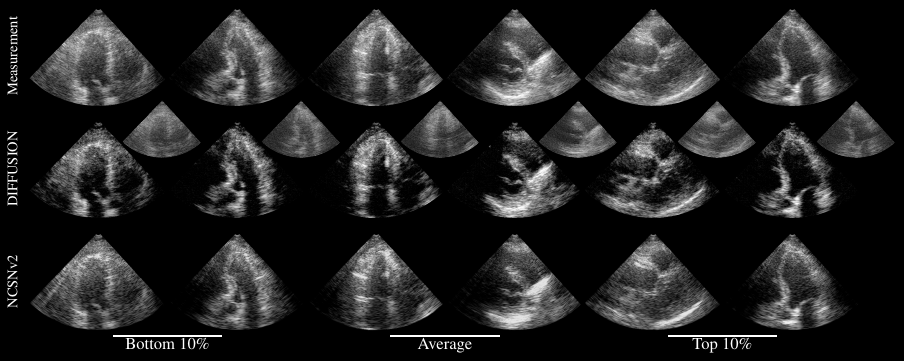}
    \caption{\emph{In-vivo} results on a random selection of the validation dataset consisting of difficult-to-image subjects. \TSW{Before selection, images were sorted based on the gCNR score of the proposed diffusion method and divided into groups related to the bottom-10 and top-10 percentile and mean (within one standard deviation). We compare resulting B-mode images of the RF-based diffusion and supervised NCSNv2 dehazing methods.} The smaller inset plots show the haze posterior samples $\bh_0$ associated with each dehazed posterior ultrasound image $\bx_0$.}
    \label{fig:comparison-invivo}
\end{figure*}

\begin{figure}
    \centering
    \includegraphics[width=\linewidth]{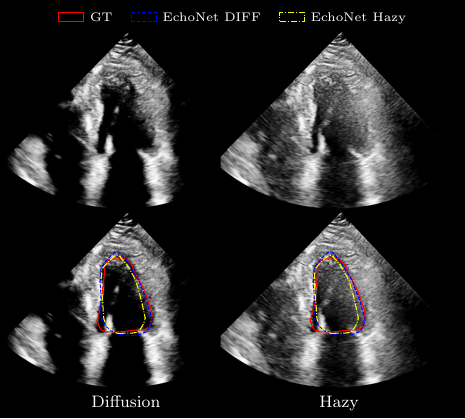}
    \caption{\TSW{Downstream segmentation task with EchoNet on both hazy input from straightforward harmonic imaging, versus the diffusion-processed images. From the estimated masks, it can be seen that the hazy input data causes EchoNet to underestimate the ventricle area (yellow) with respect to the diffusion result (blue), which delineation is more in line with the ground truth (red).}}
    \label{fig:echonet}
    \vspace{-0.5cm}
\end{figure}

A qualitative comparison is made in Fig.~\ref{fig:comparison-invivo}, where a representative set (bottom and top 10th percentile and mean gCNR scores) of the dehazed B-mode images is shown. A wider range of examples, including full cine-loops can be found on the online repository. From Fig.~\ref{fig:comparison-invivo} we observe that our diffusion method is more effective in removing haze compared to the supervised method. More importantly, apart from improved contrast, the proposed diffusion method retains and in some cases even reveals low-level tissue. Even though the \emph{in-vivo} data is out-of-distribution for the NCSNv2 model, the network can produce decent reconstructions due to the patch-based approach, showcasing the robustness of patch-based inference. Additionally, we show the posterior haze samples from the diffusion dehazing method, which resemble the haze samples used for training. We moreover observe that we can efficiently fine-tune the amount of dehazing by controlling the haze amplitude parameter $\gamma$ in the data-consistency step of our diffusion model. Note that this tunability is an inherent feature of the proposed reverse diffusion scheme with explicit data-consistency steps, and e.g. requires no gamma-specific training, in contrast to supervised methods, e.g. NCSNv2. Fig.~\ref{fig:varying_gamma} shows a comparison for varying $\gamma$-settings. The ability to fine-tune the algorithm's aggressiveness is highly beneficial, as optimal image quality is often subject to the personal preference of the user, which can significantly vary in clinical practice. Furthermore, sweeping the dehazing intensity allows the user to assess the quality of the output and set the optimal trade-off point with reduced haze and minimal signal loss, with the latter being crucial for diagnostic confidence. Automatic preference-based tuning of this parameter is an avenue for future work. Lastly, we assess the performance of the proposed diffusion method using a downstream delineation task as outlined in Section~\ref{sec:downstream} and observe that the diffusion-based processing aids EchoNet in producing more accurate delineations (Dice=0.891) compared to the hazy original image (Dice=0.880), as seen in Fig.~\ref{fig:echonet}. The Dice coefficients are averaged over the entire test set and computed with respect to the ground truth masks.


\section{Discussion}
\color{black}
\label{sec:discussion}
In this study, we developed a dehazing method for cardiac ultrasound imaging using a new joint posterior reverse diffusion scheme. Additionally, we propose various techniques that allow the learning and deployment of ultrasound priors in the RF domain. We show the benefits of RF-based dehazing over dehazing in the image domain. The proposed method outperforms a competitive baseline (BM3D on RF data) across \emph{in-vitro} and \emph{in-vivo} experiments, both quantitatively and qualitatively. We also observe that RF-domain diffusion dehazing outperforms the baselines, by preserving weakly reflected tissue signals, improving contrast, and retaining lateral resolution. Moreover, our patch-based inference scheme enables artifact-free processing of an arbitrary amount of ultrasound data and possibly can be extended to 3D ultrasound. \TSW{Even though the proposed method shows robustness in the phantom to \emph{in-vivo} experiments, it would be valuable to assess the method's performance across different ultrasound scanners, settings, and a more diverse patient cohort, providing a broader understanding of its adaptability and generalizability.}

The proposed technique effectively separates tissue and haze based on individual frames. In the future, we plan to extend this by utilizing the slow-time axis. Knowledge of past frames can potentially enhance the dehazing performance, as haze characteristics can be distinguished from tissue characteristics over time. Moreover, incorporating previous solutions has the potential to expedite the inference process, as subsequent frames are highly correlated and the current solution can be used as initialization of the next frame. However, this aspect requires further research.

Although not the focus of this work, the diffusion dehazing method is currently not ready for real-time, low-latency, deployment. The iterative nature of the sampling process combined with the necessity of two independent score-based networks results in a high number of neural function evaluations equivalent to $2\times T \times \tau$. Therefore, the method is currently intended for offline processing only. However, the fast pace of advancements in diffusion models provides a promising outlook on substantial speed-up of the reverse diffusion process \cite{jingSubspaceDiffusionGenerative2022}. Given the flexibility of the proposed method, improved sampling schemes are easily adopted, potentially enabling real-time implementation in the future.

Another promising avenue for future exploration involves the substitution of the standard Gaussian corruption process with arbitrary degradation, as proposed by Bansal~\emph{et al.} \cite{bansal2022cold}. Despite the lack of a firmly established theoretical framework of their method for incorporating arbitrary noise, we find merit in exploring this approach, especially within the context of our work on dehazing. Specifically, the idea of utilizing inherent haze as destructive noise in the reverse diffusion process is a compelling strategy. However, more research into the principles and effectiveness of this alternative approach is needed to better understand its potential for ultrasound and medical imaging as a whole.

More broadly, learning priors using deep generative models in the RF domain as proposed in this work could find practical use in various ultrasound applications, including aberration correction and adaptive beamforming. By incorporating deep priors at the beginning of the signal chain, these procedures can take advantage of the richer nature of RF signals compared to envelope-detected image data.

Lastly, we can exploit the data consistency weighting terms for more fine-grained (or even adaptive) control of the amount of dehazing. For instance, initial haze level estimations can be used to construct depth-varying data consistency weighting. Unlike discriminative or supervised methods, this provides an easy way to tune the method to taste as it requires no retraining of the networks. We leave the exploration of the discussed improvements for the proposed method for future work.

\section{Conclusions}
\label{sec:conclusions} 
To this day, many ultrasound cardiac exams are rendered unusable due to severe degradation in image quality often caused by haze. We show that existing techniques, such as harmonic imaging, are incapable of delivering high-quality B-mode imaging on difficult-to-image patients. In this work, we propose an effective dehazing framework using diffusion models that fully exploits our knowledge of the haze distribution. We present ways to efficiently learn ultrasound priors in the RF domain using deep generative modeling. Even though this work specifically focuses on the dehazing problem, the proposed posterior sampling method can be applied to a wider range of inverse problems in ultrasound. Ultimately, our method is able to achieve improved contrast on the \emph{in-vivo} cardiac data whilst preserving speckle statistics and lateral resolution. Future work should investigate ways of improving the computational efficiency of the method for real-time deployment.

\bibliography{references}

\begin{thebibliography}{10}
\providecommand{\url}[1]{#1}
\csname url@samestyle\endcsname
\providecommand{\newblock}{\relax}
\providecommand{\bibinfo}[2]{#2}
\providecommand{\BIBentrySTDinterwordspacing}{\spaceskip=0pt\relax}
\providecommand{\BIBentryALTinterwordstretchfactor}{4}
\providecommand{\BIBentryALTinterwordspacing}{\spaceskip=\fontdimen2\font plus
\BIBentryALTinterwordstretchfactor\fontdimen3\font minus
  \fontdimen4\font\relax}
\providecommand{\BIBforeignlanguage}[2]{{%
\expandafter\ifx\csname l@#1\endcsname\relax
\typeout{** WARNING: IEEEtran.bst: No hyphenation pattern has been}%
\typeout{** loaded for the language `#1'. Using the pattern for}%
\typeout{** the default language instead.}%
\else
\language=\csname l@#1\endcsname
\fi
#2}}
\providecommand{\BIBdecl}{\relax}
\BIBdecl

\bibitem{daveRecentTechnologicalAdvancements2018}
J.~K. Dave, M.~E. Mc~Donald, P.~Mehrotra, A.~R. Kohut, J.~R. Eisenbrey, and
  F.~Forsberg, ``Recent technological advancements in cardiac ultrasound
  imaging,'' \emph{Ultrasonics}, vol.~84, pp. 329--340, 2018.

\bibitem{fatemiStudyingOriginReverberation2019}
A.~Fatemi, E.~A.~R. Berg, and A.~{Rodriguez-Molares}, ``Studying the {{Origin}}
  of {{Reverberation Clutter}} in {{Echocardiography}}: {{In Vitro
  Experiments}} and {{In Vivo Demonstrations}},'' \emph{Ultrasound in Medicine
  \& Biology}, vol.~45, no.~7, pp. 1799--1813, July 2019.

\bibitem{uppotImpactObesityRadiology2007}
R.~N. Uppot, ``Impact of {{Obesity}} on {{Radiology}},'' \emph{Radiologic
  Clinics of North America}, vol.~45, no.~2, pp. 231--246, March 2007.

\bibitem{sassaroliImageQualityEvaluation2019}
E.~Sassaroli, C.~Crake, A.~Scorza, D.-S. Kim, and M.-A. Park, ``Image quality
  evaluation of ultrasound imaging systems: Advanced {{B-modes}},''
  \emph{Journal of Applied Clinical Medical Physics}, vol.~20, no.~3, pp.
  115--124, 2019.

\bibitem{chenDeepLearningCardiac2020}
C.~Chen, C.~Qin, H.~Qiu, G.~Tarroni, J.~Duan, W.~Bai, and D.~Rueckert, ``Deep
  {{Learning}} for {{Cardiac Image Segmentation}}: {{A Review}},''
  \emph{Frontiers in Cardiovascular Medicine}, vol.~7, p.~25, March 2020.

\bibitem{tranquartClinicalUseUltrasound1999}
F.~Tranquart, N.~Grenier, V.~Eder, and L.~Pourcelot, ``Clinical use of
  ultrasound tissue harmonic imaging,'' \emph{Ultrasound in Medicine \&
  Biology}, vol.~25, no.~6, pp. 889--894, July 1999.

\bibitem{wellsUltrasoundImaging2006}
P.~N.~T. Wells, ``Ultrasound imaging,'' \emph{Physics in Medicine \& Biology},
  vol.~51, no.~13, p. R83, June 2006.

\bibitem{yangDAGANDeepDeAliasing2018}
G.~Yang, S.~Yu, H.~Dong, G.~Slabaugh, P.~L. Dragotti, X.~Ye, F.~Liu,
  S.~Arridge, J.~Keegan, Y.~Guo, and D.~Firmin, ``{{DAGAN}}: {{Deep De-Aliasing
  Generative Adversarial Networks}} for {{Fast Compressed Sensing MRI
  Reconstruction}},'' \emph{IEEE Transactions on Medical Imaging}, vol.~37,
  no.~6, pp. 1310--1321, June 2018.

\bibitem{youCTSuperResolutionGAN2020}
C.~You, G.~Li, Y.~Zhang, X.~Zhang, H.~Shan, M.~Li, S.~Ju, Z.~Zhao, Z.~Zhang,
  W.~Cong, M.~W. Vannier, P.~K. Saha, E.~A. Hoffman, and G.~Wang, ``{{CT
  Super-Resolution GAN Constrained}} by the {{Identical}}, {{Residual}}, and
  {{Cycle Learning Ensemble}} ({{GAN-CIRCLE}}),'' \emph{IEEE Transactions on
  Medical Imaging}, vol.~39, no.~1, pp. 188--203, January 2020.

\bibitem{songSolvingInverseProblems2022}
\BIBentryALTinterwordspacing
Y.~Song, L.~Shen, L.~Xing, and S.~Ermon, ``Solving inverse problems in medical
  imaging with score-based generative models,'' in \emph{International
  Conference on Learning Representations}, 2022. [Online]. Available:
  \url{https://openreview.net/forum?id=vaRCHVj0uGI}
\BIBentrySTDinterwordspacing

\bibitem{vandeschaftUltrasoundSpeckleSuppression2021}
V.~van~de Schaft and R.~J.~G. van Sloun, ``Ultrasound speckle suppression and
  denoising using mri-derived normalizing flow priors,'' 2021.

\bibitem{chungScorebasedDiffusionModels2022}
H.~Chung and J.~C. Ye, ``Score-based diffusion models for accelerated
  {{MRI}},'' \emph{Medical Image Analysis}, vol.~80, p. 102479, August 2022.

\bibitem{vanslounDeepLearningUltrasound2020}
R.~J.~G. {van Sloun}, R.~Cohen, and Y.~C. Eldar, ``Deep {{Learning}} in
  {{Ultrasound Imaging}},'' \emph{Proceedings of the IEEE}, vol. 108, no.~1,
  pp. 11--29, January 2020.

\bibitem{zhangDeepStableLearning2021}
X.~Zhang, P.~Cui, R.~Xu, L.~Zhou, Y.~He, and Z.~Shen, ``Deep {{Stable
  Learning}} for {{Out-Of-Distribution Generalization}},'' April 2021.

\bibitem{chatterjeeGeneralizationMysteryDeep2022}
S.~Chatterjee and P.~Zielinski, ``On the generalization mystery in deep
  learning,'' 2022.

\bibitem{kingma2018glow}
D.~P. Kingma and P.~Dhariwal, ``Glow: Generative flow with invertible 1x1
  convolutions,'' \emph{Advances in neural information processing systems},
  vol.~31, 2018.

\bibitem{dabovImageDenoisingBlockmatching2006}
K.~Dabov, A.~Foi, V.~Katkovnik, and K.~Egiazarian, ``Image denoising with
  block-matching and {{3D}} filtering,'' in \emph{Image {{Processing}}:
  {{Algorithms}} and {{Systems}}, {{Neural Networks}}, and {{Machine
  Learning}}}, vol. 6064.\hskip 1em plus 0.5em minus 0.4em\relax {SPIE},
  February 2006, pp. 354--365.

\bibitem{santosUltrasoundImageDespeckling2017}
C.~A.~N. Santos, D.~L.~N. Martins, and N.~D.~A. Mascarenhas, ``Ultrasound
  {{Image Despeckling Using Stochastic Distance-Based BM3D}},'' \emph{IEEE
  Transactions on Image Processing}, vol.~26, no.~6, pp. 2632--2643, June 2017.

\bibitem{yueNonlinearMultiscaleWavelet2006}
Y.~Yue, M.~Croitoru, A.~Bidani, J.~Zwischenberger, and J.~Clark, ``Nonlinear
  multiscale wavelet diffusion for speckle suppression and edge enhancement in
  ultrasound images,'' \emph{IEEE Transactions on Medical Imaging}, vol.~25,
  no.~3, pp. 297--311, March 2006.

\bibitem{geyerAssessmentMyocardialMechanics2010}
H.~Geyer, G.~Caracciolo, H.~Abe, S.~Wilansky, S.~Carerj, F.~Gentile, H.-J.
  Nesser, B.~Khandheria, J.~Narula, and P.~P. Sengupta, ``Assessment of
  {{Myocardial Mechanics Using Speckle Tracking Echocardiography}}:
  {{Fundamentals}} and {{Clinical Applications}},'' \emph{Journal of the
  American Society of Echocardiography}, vol.~23, no.~4, pp. 351--369, April
  2010.

\bibitem{sjoerdsmaSpatialNearFieldClutter2021}
M.~Sjoerdsma, S.~Bouwmeester, P.~Houthuizen, F.~N. {van de Vosse}, and R.~G.~P.
  Lopata, ``A {{Spatial Near-Field Clutter Reduction Filter Preserving Tissue
  Speckle}} in {{Echocardiography}},'' \emph{IEEE Transactions on Ultrasonics,
  Ferroelectrics, and Frequency Control}, vol.~68, no.~4, pp. 979--992, April
  2021.

\bibitem{turekClutterMitigationEchocardiography2015}
J.~S. Turek, M.~Elad, and I.~Yavneh, ``Clutter {{Mitigation}} in
  {{Echocardiography Using Sparse Signal Separation}},'' \emph{International
  Journal of Biomedical Imaging}, vol. 2015, pp. 1--18, 2015.

\bibitem{solomonDeepUnfoldedRobust2020}
O.~Solomon, R.~Cohen, Y.~Zhang, Y.~Yang, Q.~He, J.~Luo, R.~J.~G. {van Sloun},
  and Y.~C. Eldar, ``Deep {{Unfolded Robust PCA With Application}} to {{Clutter
  Suppression}} in {{Ultrasound}},'' \emph{IEEE Transactions on Medical
  Imaging}, vol.~39, no.~4, pp. 1051--1063, April 2020.

\bibitem{wildeboerBlindSourceSeparation2020}
R.~R. Wildeboer, F.~Sammali, R.~J.~G. {van Sloun}, Y.~Huang, P.~Chen, M.~Bruce,
  C.~Rabotti, S.~Shulepov, G.~Salomon, B.~C. Schoot, H.~Wijkstra, and
  M.~Mischi, ``Blind {{Source Separation}} for {{Clutter}} and {{Noise
  Suppression}} in {{Ultrasound Imaging}}: {{Review}} for {{Different
  Applications}},'' \emph{IEEE Transactions on Ultrasonics, Ferroelectrics, and
  Frequency Control}, vol.~67, no.~8, pp. 1497--1512, August 2020.

\bibitem{liuDeepLearningMedical2019}
S.~Liu, Y.~Wang, X.~Yang, B.~Lei, L.~Liu, S.~X. Li, D.~Ni, and T.~Wang, ``Deep
  {{Learning}} in {{Medical Ultrasound Analysis}}: {{A Review}},''
  \emph{Engineering}, vol.~5, no.~2, pp. 261--275, April 2019.

\bibitem{vignonResolutionImprovementFully2020}
F.~Vignon, J.~S. Shin, F.~C. Meral, I.~Apostolakis, S.-W. Huang, and J.-L.
  Robert, ``Resolution {{Improvement}} with a {{Fully Convolutional Neural
  Network Applied}} to {{Aligned Per-Channel}} data,'' in \emph{2020 {{IEEE
  International Ultrasonics Symposium}} ({{IUS}})}, September 2020, pp. 1--4.

\bibitem{stevensAcceleratedIntravascularUltrasound2022}
T.~S. Stevens, N.~Chennakeshava, F.~J. {de Bruijn}, M.~Peka{\v r}, and R.~J.
  {van Sloun}, ``Accelerated {{Intravascular Ultrasound Imaging}} using {{Deep
  Reinforcement Learning}},'' in \emph{{{ICASSP}} 2022 - 2022 {{IEEE
  International Conference}} on {{Acoustics}}, {{Speech}} and {{Signal
  Processing}} ({{ICASSP}})}, May 2022, pp. 1216--1220.

\bibitem{chennakeshavaDeepProximalUnfolding2022}
N.~Chennakeshava, T.~S. Stevens, F.~J. {de Bruijn}, A.~Hancock, M.~Peka{\v r},
  Y.~C. Eldar, M.~Mischi, and R.~J. {van Sloun}, ``Deep {{Proximal Unfolding
  For Image Recovery}} from {{Under-Sampled Channel Data}} in {{Intravascular
  Ultrasound}},'' in \emph{{{ICASSP}} 2022 - 2022 {{IEEE International
  Conference}} on {{Acoustics}}, {{Speech}} and {{Signal Processing}}
  ({{ICASSP}})}, May 2022, pp. 1221--1225.

\bibitem{luijtenUltrasoundSignalProcessing2023}
B.~Luijten, N.~Chennakeshava, Y.~C. Eldar, M.~Mischi, and R.~J.~G. {van Sloun},
  ``Ultrasound {{Signal Processing}}: {{From Models}} to {{Deep Learning}},''
  \emph{Ultrasound in Medicine \& Biology}, vol.~49, no.~3, pp. 677--698, 2023.

\bibitem{jahrenReverberationSuppressionEchocardiography2023}
T.~S. Jahren, A.~R. S{\o}rnes, B.~D{\'e}nari{\'e}, E.~Steen, T.~Bj{\aa}stad,
  and A.~H.~S. Solberg, ``Reverberation {{Suppression}} in {{Echocardiography
  Using}} a {{Causal Convolutional Neural Network}},'' \emph{IEEE Access},
  vol.~11, pp. 67\,922--67\,937, 2023.

\bibitem{asgariandehkordiDeepUltrasoundDenoising2023}
H.~Asgariandehkordi, S.~Goudarzi, A.~Basarab, and H.~Rivaz, ``Deep ultrasound
  denoising using diffusion probabilistic models,'' 2023.

\bibitem{stojanovskiEchoNoiseSynthetic2023}
D.~Stojanovski, U.~Hermida, P.~Lamata, A.~Beqiri, and A.~Gomez, ``Echo from
  noise: synthetic ultrasound image generation using diffusion models for real
  image segmentation,'' 2023.

\bibitem{kazerouniDiffusionModelsMedical2023}
A.~Kazerouni, E.~K. Aghdam, M.~Heidari, R.~Azad, M.~Fayyaz, I.~Hacihaliloglu,
  and D.~Merhof, ``Diffusion {{Models}} for {{Medical Image Analysis}}: {{A
  Comprehensive Survey}},'' June 2023.

\bibitem{hoDenoisingDiffusionProbabilistic2020}
J.~Ho, A.~Jain, and P.~Abbeel, ``Denoising {{Diffusion Probabilistic
  Models}},'' in \emph{Advances in {{Neural Information Processing Systems}}},
  vol.~33.\hskip 1em plus 0.5em minus 0.4em\relax {Curran Associates, Inc.},
  2020, pp. 6840--6851.

\bibitem{songScoreBasedGenerativeModeling2021}
\BIBentryALTinterwordspacing
Y.~Song, J.~Sohl-Dickstein, D.~P. Kingma, A.~Kumar, S.~Ermon, and B.~Poole,
  ``Score-based generative modeling through stochastic differential
  equations,'' in \emph{International Conference on Learning Representations},
  2021. [Online]. Available: \url{https://openreview.net/forum?id=PxTIG12RRHS}
\BIBentrySTDinterwordspacing

\bibitem{karrasElucidatingDesignSpace2022}
\BIBentryALTinterwordspacing
T.~Karras, M.~Aittala, T.~Aila, and S.~Laine, ``Elucidating the design space of
  diffusion-based generative models,'' in \emph{Advances in Neural Information
  Processing Systems}, S.~Koyejo, S.~Mohamed, A.~Agarwal, D.~Belgrave, K.~Cho,
  and A.~Oh, Eds., vol.~35.\hskip 1em plus 0.5em minus 0.4em\relax Curran
  Associates, Inc., 2022, pp. 26\,565--26\,577. [Online]. Available:
  \url{https://proceedings.neurips.cc/paper_files/paper/2022/file/a98846e9d9cc01cfb87eb694d946ce6b-Paper-Conference.pdf}
\BIBentrySTDinterwordspacing

\bibitem{vincentConnectionScoreMatching2011}
P.~Vincent, ``A {{Connection Between Score Matching}} and {{Denoising
  Autoencoders}},'' \emph{Neural Computation}, vol.~23, no.~7, pp. 1661--1674,
  July 2011.

\bibitem{songGenerativeModelingEstimating2019}
Y.~Song and S.~Ermon, ``Generative {{Modeling}} by {{Estimating Gradients}} of
  the {{Data Distribution}},'' in \emph{Advances in {{Neural Information
  Processing Systems}}}, vol.~32.\hskip 1em plus 0.5em minus 0.4em\relax
  {Curran Associates, Inc.}, 2019.

\bibitem{stevensRemovingStructuredNoise2023}
T.~S.~W. Stevens, H.~van Gorp, F.~C. Meral, J.~Shin, J.~Yu, J.-L. Robert, and
  R.~J.~G. van Sloun, ``Removing structured noise with diffusion models,''
  2023.

\bibitem{chungDiffusionPosteriorSampling2023}
\BIBentryALTinterwordspacing
H.~Chung, J.~Kim, M.~T. Mccann, M.~L. Klasky, and J.~C. Ye, ``Diffusion
  posterior sampling for general noisy inverse problems,'' in \emph{The
  Eleventh International Conference on Learning Representations}, 2023.
  [Online]. Available: \url{https://openreview.net/forum?id=OnD9zGAGT0k}
\BIBentrySTDinterwordspacing

\bibitem{mengDiffusionModelBased2023}
X.~Meng and Y.~Kabashima, ``Diffusion model based posterior sampling for noisy
  linear inverse problems,'' 2024.

\bibitem{fengScoreBasedDiffusionModels2023}
B.~T. Feng, J.~Smith, M.~Rubinstein, H.~Chang, K.~L. Bouman, and W.~T. Freeman,
  ``Score-based diffusion models as principled priors for inverse imaging,''
  2023.

\bibitem{sklar2021digital}
B.~Sklar, \emph{Digital communications: fundamentals and applications}.\hskip
  1em plus 0.5em minus 0.4em\relax Pearson, 2021.

\bibitem{koutiniReceptiveFieldRegularization2021}
K.~Koutini, H.~{Eghbal-zadeh}, and G.~Widmer, ``Receptive {{Field
  Regularization Techniques}} for {{Audio Classification}} and {{Tagging With
  Deep Convolutional Neural Networks}},'' \emph{IEEE/ACM Transactions on Audio,
  Speech, and Language Processing}, vol.~29, pp. 1987--2000, 2021.

\bibitem{zhangDiffCollageParallelGeneration2023}
Q.~Zhang, J.~Song, X.~Huang, Y.~Chen, and M.-Y. Liu, ``Diffcollage: Parallel
  generation of large content with diffusion models,'' in \emph{Proceedings of
  the IEEE/CVF Conference on Computer Vision and Pattern Recognition (CVPR)},
  June 2023, pp. 10\,188--10\,198.

\bibitem{wangZeroShotImageRestoration2022}
\BIBentryALTinterwordspacing
Y.~Wang, J.~Yu, and J.~Zhang, ``Zero-shot image restoration using denoising
  diffusion null-space model,'' in \emph{The Eleventh International Conference
  on Learning Representations}, 2023. [Online]. Available:
  \url{https://openreview.net/forum?id=mRieQgMtNTQ}
\BIBentrySTDinterwordspacing

\bibitem{sahariaPaletteImagetoImageDiffusion2022}
C.~Saharia, W.~Chan, H.~Chang, C.~Lee, J.~Ho, T.~Salimans, D.~Fleet, and
  M.~Norouzi, ``Palette: {{Image-to-Image Diffusion Models}},'' in
  \emph{Special {{Interest Group}} on {{Computer Graphics}} and {{Interactive
  Techniques Conference Proceedings}}}.\hskip 1em plus 0.5em minus 0.4em\relax
  {Vancouver BC Canada}: {ACM}, August 2022.

\bibitem{vignonRevisitingWienerPostfilter2020}
F.~Vignon, A.~Sadeghi, J.~Yu, F.~C. Meral, I.~Apostolakis, J.~S. Shin, and
  J.-L. Robert, ``Revisiting the {{Wiener}} postfilter for ultrasound image
  quality improvement,'' in \emph{2020 {{IEEE International Ultrasonics
  Symposium}} ({{IUS}})}, September 2020, pp. 1--4.

\bibitem{chungComeCloserDiffuseFasterAcceleratingConditional2022}
H.~Chung, B.~Sim, and J.~C. Ye, ``Come-{{Closer-Diffuse-Faster}}:
  {{Accelerating Conditional Diffusion Models}} for {{Inverse Problems}}
  through {{Stochastic Contraction}},'' in \emph{2022 {{IEEE}}/{{CVF
  Conference}} on {{Computer Vision}} and {{Pattern Recognition}} ({{CVPR}})},
  June 2022, pp. 12\,403--12\,412.

\bibitem{songImprovedTechniquesTraining2020}
Y.~Song and S.~Ermon, ``Improved {{Techniques}} for {{Training Score-Based
  Generative Models}},'' in \emph{Advances in {{Neural Information Processing
  Systems}}}, vol.~33.\hskip 1em plus 0.5em minus 0.4em\relax {Curran
  Associates, Inc.}, 2020, pp. 12\,438--12\,448.

\bibitem{rodriguez-molaresGeneralizedContrasttonoiseRatio2020}
A.~{Rodriguez-Molares}, O.~M.~H. Rindal, J.~D'hooge, v.-E. M{\aa}s{\o}y,
  A.~Austeng, M.~A.~L. Bell, and H.~Torp, ``The generalized contrast-to-noise
  ratio: A formal definition for lesion detectability,'' \emph{IEEE
  transactions on ultrasonics, ferroelectrics, and frequency control}, vol.~67,
  no.~4, pp. 745--759, April 2020.

\bibitem{moal2022explicit}
O.~Moal, E.~Roger, A.~Lamouroux, C.~Younes, G.~Bonnet, B.~Moal, and S.~Lafitte,
  ``Explicit and automatic ejection fraction assessment on 2d cardiac
  ultrasound with a deep learning-based approach,'' \emph{Computers in biology
  and medicine}, vol. 146, p. 105637, 2022.

\bibitem{ouyangVideobasedAIBeattobeat2020}
D.~Ouyang, B.~He, A.~Ghorbani, N.~Yuan, J.~Ebinger, C.~P. Langlotz, P.~A.
  Heidenreich, R.~A. Harrington, D.~H. Liang, E.~A. Ashley, and J.~Y. Zou,
  ``Video-based {{AI}} for beat-to-beat assessment of cardiac function,''
  \emph{Nature}, vol. 580, no. 7802, pp. 252--256, April 2020.

\bibitem{jingSubspaceDiffusionGenerative2022}
B.~Jing, G.~Corso, R.~Berlinghieri, and T.~Jaakkola, ``Subspace diffusion
  generative models,'' in \emph{European Conference on Computer Vision}.\hskip
  1em plus 0.5em minus 0.4em\relax Springer, 2022, pp. 274--289.

\bibitem{bansal2022cold}
A.~Bansal, E.~Borgnia, H.-M. Chu, J.~S. Li, H.~Kazemi, F.~Huang, M.~Goldblum,
  J.~Geiping, and T.~Goldstein, ``Cold diffusion: Inverting arbitrary image
  transforms without noise,'' \emph{arXiv preprint arXiv:2208.09392}, 2022.

\end{thebibliography}

\end{document}